\documentclass[11pt]{article}

\usepackage{amssymb,epsfig,fullpage}

\begin{document}

\title{Amplitude and phase dynamics in oscillators with distributed-delay coupling}

\author{Y.N. Kyrychko\thanks{Corresponding author. Email: y.kyrychko@sussex.ac.uk}, \hspace{0.5cm}K.B. Blyuss 
\\\\ Department of Mathematics, University of Sussex, Falmer,\\
Brighton, BN1 9QH, United Kingdom\\\\
\and E. Sch\"oll
\\\\ Institut f\"ur Theoretische Physik, Technische Universit\"at Berlin,\\
10623 Berlin, Germany}

\maketitle

\begin{abstract}
This paper studies the effects of distributed delay coupling on the dynamics in a system of
non-identical coupled Stuart-Landau oscillators. For uniform and gamma delay distribution
kernels, conditions for amplitude death are obtained in terms of average frequency, frequency
detuning and parameters of the coupling, including coupling strength and phase, as well
as the mean time delay and the width of the delay distribution. To gain further insight into the
dynamics inside amplitude death regions, eigenvalues of the corresponding characteristic
equations are computed numerically. Oscillatory dynamics of the system is also investigated
using amplitude and phase representation. Various branches of phase-locked solutions are
identified, and their stability is analysed for different types of delay distributions.
\end{abstract}

\section{Introduction}

Dynamics of many complex physical, biological, and engineering systems can be effectively modelled mathematically using ensembles of coupled oscillators (Kuramoto 1984, Pikovsky {\it et al.} 2001). A significant advantage of such an approach over other methodologies lies in the possibility to identify and study regimes with particular types of behaviour, such as emergence and stability of cooperation and (de)synchronization, different kinds of spatial patterns, travelling waves etc. Such analyses often provide very important insights for practical applications. Our understanding and treatment of various brain pathologies and deficiencies, such as Parkinson's, Alzheimer's, epilepsy, heavily relies on the analysis of synchronization of oscillating neural populations (Uhlhaas \& Singer 2006, Popovych {\it et al.} 2005, Schnitzler \& Gross 2005). Recent work on laser communication networks has extensively used coupled oscillator models to study the dynamics of in-phase or anti-phase and complete chaos synchronization (Heil {\it et al.} 2001, Flunkert {\it et al.} 2009, Hicke {\it et al.} 2011, Flunkert \& Sch\"oll 2012, Kozyreff {\it et al.} 2000). 
Chimera states, where a network of oscillators splits into coexisting domains of coherent, phase-locked and incoherent, desynchronized behaviour, have also been observed (Abrams \& Strogatz 2004, Hagerstrom {\it et al.} 2012, Omelchenko {\it et al.} 2011, Kuramoto \& Battogtokh 2002, Tinsley {\it et al.} 2012).
These studies have shown that coupling between different elements can play a dual role in the dynamics: it can both lead to suppression of oscillations, and it can also facilitate a certain degree of synchronization between different elements.

When analysing the dynamics of coupled systems, it is important to take into account the fact that in many cases the coupling between different elements is not instantaneous. Time delays associated with such coupling can have a major impact on the overall dynamics of the system. In practical examples these time delays are often associated with delays in propagation or processing of signals, response time of mechanical actuators, translation and transcription time in genetic oscillators (Atay 2010, Just {\it et al.} 2010, Kyrychko \& Hogan 2010, Sch\"oll \& Schuster 2008). In all these examples, explicit inclusion of time delays in the model has provided a more realistic and accurate representation of the system under consideration. From a mathematical perspective, differential equations with time delays are infinite-dimensional dynamical systems, which makes their analysis, as well as numerical simulations, quite involved (Diekmann {\it et al.} 1995, Erneux 2009).

In the case when the coupling between oscillators is sufficiently weak, it is possible to reduce the full system to a phase-only model (Kuramoto 1984). Significant amount of research has been done over the years on the analysis of phase oscillators with different kinds of time-delayed coupling (D'Huys {\it et al.} 2008, Earl \& Strogatz 2003, Erneux \& Glorieux 2010, Nakamura 1994, Niebur 1991, Pikovsky {\it et al.} 2001, Popovych {\it et al.} 2010, Schuster \& Wagner 1989, Sen {\it et al.} 2010). Such systems of phase oscillators demonstrate a rich variety of dynamical regimes, including chaos, synchronization, splay and chimera states, characterized by co-existence of coherent and incoherent states. In many cases, coupling in the phase models leads to phase entrainment and the emergence of so-called phase-locked solutions, where all oscillators start to oscillate with the same common frequency and have a constant shift between their phases. Existence and stability of such phase-locked solutions have been studied in a number of systems of oscillators with different kinds of local and non-local coupling (Kuramoto \& Battogtokh 2002, Sen {\it et al.} 2010).

For sufficiently strong coupling between oscillators, one can observe another interesting and important aspect of dynamics of coupled oscillator systems: the ability of external coupling to suppress otherwise stable periodic oscillations. This was first discovered by Bar-Eli in the context of chemical reactions with instantaneous coupling (Bar-Eli 1985), and it has been shown for a system of two coupled Stuart-Landau oscillators that the `Bar-Eli effect' can only occur, provided the coupling strength and the frequency detuning of the two oscillators are both sufficiently large (Aronson {\it et al.} 1990, Ermentrout 1990, Mirollo \& Strogatz 1990). Later it was discovered that when the coupling is time-delayed, it is possible to achieve suppression of oscillations and stabilization of an unstable fixed point even for identical oscillators (Ramana Reddy {\it et al.} 1998,1999). This phenomenon, named amplitude death (Ramana Reddy {\it et al.} 1998), oscillator death, or 'death by delay' (Strogatz 1998), has been demonstrated experimentally in nonlinear electronic circuits (Ramana Reddy 2000), in the dynamics of slime mold {\it Physarum polycephalum} (Herrero {\it et al.} 2000) and thermo-optical oscillators linearly coupled by heat transfer (Takamatsu {\it et al.} 2000). Amplitude death has been subsequently studied for a number of different systems and couplings (Ramana Dodla {\it et al.} 2004,  Karnatak {\it et al.} 2010, Sen {\it et al.} 2005, 2010, Choe {\it et al.} 2007, 
Sch{\"o}ll {\it et al.} 2009, Zuo {\it et al.} 2011, 2012, B{\"a}r {\it et al.} 2012).

Despite successes in the analysis of systems of coupled oscillators with time-delayed coupling, one of the limitations of the majority of this research has been the restriction on the type of time-delayed coupling, which is usually taken to be in the form of one or several constant time delays. At the same time, in many realistic systems time delays themselves are not constant (Gjurchinovski \& Urumov 2008, 2010) and may either vary depending on the values of system variables (state-dependent delays) or just not be explicitly known. In order to account for such situations mathematically, one can use the formalism of distributed time delays, where the time delay is represented through an integral kernel describing a particular delay distribution (Bernard 2001, Campbell \& Jessop 2009). Distributed time delay has been successfully used to describe situations when only an approximate value of time delay is known in engineering experiments (Kiss \& Krauskopf 2011, Michiels {\it et al.} 2005), for modelling distributions of waiting times in epidemiological models (Blyuss \& Kyrychko 2010), maturation periods in population and ecological models (Faria 2010, Gourley 2003), as well as in model of traffic dynamics (Sipahi {\it et al.} 2008), neural systems (Eurich {\it et al.} 2005), predator-prey and food webs (Thiel {\it et al.} 2003).

In this paper we investigate amplitude death and phase dynamics in a system of coupled Stuart-Landau oscillators with distributed delay coupling. This system is a prototype for dynamics near a supercritical Hopf bifurcation, and in this capacity it captures essential features of many realistic systems in such a regime. The corresponding mathematical model can be written in the form
\begin{equation}\label{SL}
\begin{array}{l}
\displaystyle{\dot{z}_1(t)=(1+i\omega_1)z_{1}(t)-|z_1(t)|^2 z_1(t)+Ke^{i\theta}\left[\int_{0}^{\infty}g(t')z_{2}(t-t')dt'-z_1(t)\right],}\\\\
\displaystyle{\dot{z}_2(t)=(1+i\omega_2)z_{2}(t)-|z_2(t)|^2 z_2(t)+Ke^{i\theta}\left[\int_{0}^{\infty}g(t')z_{1}(t-t')dt'-z_2(t)\right],}
\end{array}
\end{equation}
where $z_{1,2}\in\mathbb{C}$, $\omega_{1,2}$ are frequencies of two oscillators, $K\in\mathbb{R}_{+}$ and $\theta\in\mathbb{R}$ are the strength and the phase of coupling, respectively, and $g(\cdot)$ is a kernel of delay distribution. The kernel $g$ is taken to be positive-definite and normalized to unity:
\[
g(u)\geq 0,\hspace{0.5cm} \int_{0}^{\infty}g(u)du=1.
\]
The case $g(u)=\delta(u)$ corresponds to instantaneous coupling $(z_2-z_1)$ describing a situation when oscillators interact without any delay. For $g(u)=\delta(u-\tau)$, the coupling takes the form of a discrete time delay: $[z_2(t-\tau)-z_1(t)]$. Atay (2003) has analysed this system for the case of zero coupling phase and a uniform delay distribution kernel analytically for the case of identical oscillators and numerically for non-identical oscillators, and identified regimes of amplitude death in terms of coupling strength and mean time delay. Kyrychko {\it et al.} (2011) have analysed amplitude death in model (\ref{SL}) with $\omega_1=\omega_2=\omega_0$ for uniform and gamma distributed delay kernels and non-zero phase.

The outline of this paper is as follows. In the next section we analyse stability of the trivial equlibrium of system (\ref{SL}) and find regions of amplitude death depending on parameters of the coupling and delay distribution. The eigenvalues of the corresponding characteristic equations determining the stability of the steady state are computed numerically to gain a better understanding of system dynamics inside amplitude death regimes. Section 3 contains analysis of phase-locked solutions of system (\ref{SL}) for uniform and gamma distributions in the case of constant equal amplitudes of oscillations (when the system reduces to a coupled system of Kuramoto oscillators with distributed-delay coupling), as well as in a general case of arbitrary amplitudes. In each case, we identify branches of phase-locked solutions and numerically compute their stability. The paper concludes with a discussion of results as well as possible further developments of this work.

\section{Amplitude death}

To study the possibility of amplitude death in the system (\ref{SL}), we linearize this system near the trivial steady state $z_{1,2}=0$. The corresponding characteristic equation is given by
\begin{equation}\label{ch_eq}
\left(1+i\omega_1 -Ke^{i\theta}-\lambda\right)\left(1+i\omega_2 -Ke^{i\theta}-\lambda\right)-K^{2}e^{2i\theta}\left[\{\mathcal{L}g\}(\lambda)\right]^{2}=0,
\end{equation}
where $\lambda$ is an eigenvalue of the Jacobian, and
\begin{equation}
\{\mathcal{L}g\}(s)=\int_{0}^{\infty}e^{-su}g(u)du,
\end{equation}
is the Laplace transform of the function $g(u)$. Atay (2003) has investigated the case $\theta=0$ analytically for the case of identical oscillators $\omega_1=\omega_2=\omega_0$, and numerically for $\omega_1\neq\omega_2$ and a uniform delay distribution kernel. More recently, Kyrychko {\it et al.} (2011) have studied analytically and numerically the case of $\omega_1=\omega_2=\omega_0$ and a non-vanishing coupling phase $\theta\neq 0$ with uniform and gamma delay distributions.

To make further analytical progress, it is instructive to specify a
particular choice of the delay kernel. As a first example, we consider a uniformly distributed kernel
\begin{equation}\label{UKer}
g(u)=\left\{
\begin{array}{l}
\displaystyle{\frac{1}{2\rho} \hspace{1cm}\mbox{for }\tau-\rho\leq u\leq \tau+\rho,}\\\\
0\hspace{1cm}\mbox{elsewhere.}
\end{array}
\right.
\end{equation}
This distribution has the mean time delay 
\[
\tau_{m}\equiv<\tau>=\int_{0}^{\infty}ug(u)du=\tau,
\]
and the variance
\begin{equation}\label{VD}
\displaystyle{\sigma^2=\int_{0}^{\infty}(u-\tau_m)^{2}g(u)du=\frac{\rho^{2}}{3}.}
\end{equation}
In the case of uniformly distributed kernel (\ref{UKer}), it is quite easy to compute the Laplace transform of the distribution $g(u)$ as:
\[
\{\mathcal{L}g\}(\lambda)=\frac{1}{2\rho\lambda}e^{-\lambda\tau}\left(e^{\lambda\rho}-e^{-\lambda\rho}\right)=e^{-\lambda\tau}
\frac{\sinh(\lambda\rho)}{\lambda\rho},
\]

and this transforms the characteristic equation (\ref{ch_eq}) into
\begin{equation}\label{ch_eq_2}
\left(1+i\omega_1 -Ke^{i\theta}-\lambda\right)\left(1+i\omega_2 -Ke^{i\theta}-\lambda\right)=K^{2}e^{2i\theta}e^{-2\lambda\tau}
\left[\frac{\sinh(\lambda\rho)}{\lambda\rho}\right]^2.
\end{equation}
Since the roots of the characteristic equation (\ref{ch_eq_2}) are complex-valued, stability of the trivial steady state can only change if some of these eigenvalues cross the
imaginary axis. To this end, we can look for characteristic roots in the form $\lambda=i\omega$. Substituting this into the characteristic equation (\ref{ch_eq_2}) and separating real and imaginary parts gives the following system of equations for $(K,\tau)$ as parameterized by the Hopf frequency $\omega$:
\begin{equation}\label{Ktau}
\begin{array}{l}
\displaystyle{(1-K\cos\theta)^2-(\overline{\omega}-\omega-K\sin\theta)^2+\frac{\Delta^2}{4}=K^2\delta(\rho,\omega)\cos[2(\theta-\omega\tau)],}\\\\
\displaystyle{2(1-K\cos\theta)(\overline{\omega}-K\sin\theta-\omega)=K^2\delta(\rho,\omega)\sin[2(\theta-\omega\tau)],}
\end{array}
\end{equation}

\newpage
\begin{figure}[h]
\epsfig{file=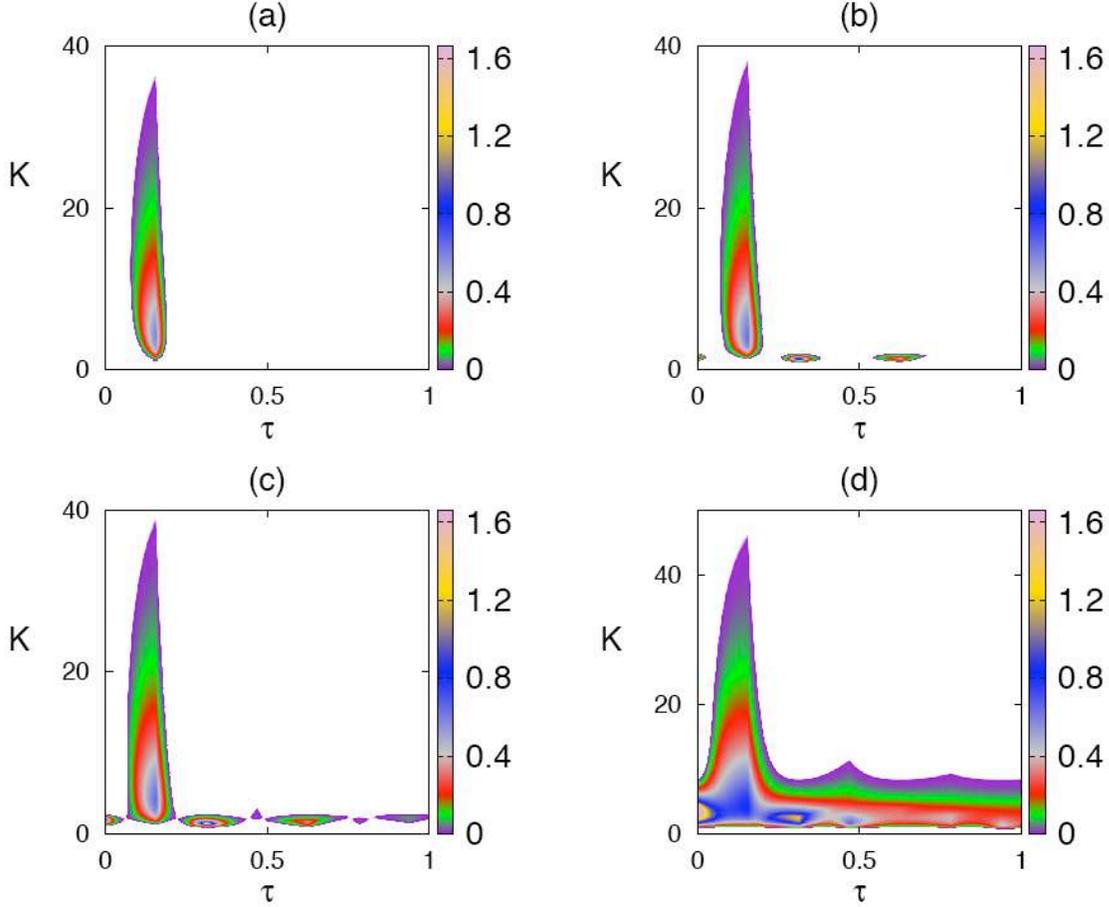,width=15cm}
\caption{(Colour online) Regimes of amplitude death depending on the coupling strength $K$ and $\tau$ for the uniform  distribution kernel with $\theta=0$, $\rho=0$ (discrete time delay), and $\overline{\omega}=10$. Colour code denotes $[-{\rm max}\{{\rm Re}(\lambda)\}]$ for ${\rm max}\{{\rm Re}(\lambda)\}\le 0$. (a) $\Delta=0$. (b) $\Delta=1.45$. (c) $\Delta=1.65$. (d) $\Delta=1.75$.}\label{uni1}
\end{figure}

\noindent where
\[
\delta(\rho,\omega)=\left[\frac{\sin(\omega\rho)}{\omega\rho}\right]^2,
\]
and we have introduced the frequency mismatch (detuning) $\Delta$ and the mean frequency 
$\overline{\omega}$ as
\[
\Delta=\omega_1-\omega_2,\hspace{0.5cm}\overline{\omega}=\frac{\omega_1+\omega_2}{2}.
\]
Squaring and adding the two equations in (\ref{Ktau}) gives a single quartic equation for the coupling strength $K$
\begin{equation}\label{Freq}
\begin{array}{l}
\displaystyle{\left[(1-K\cos\theta)^2-(\overline{\omega}-\omega-K\sin\theta)^2+\frac{\Delta^2}{4}\right]^2}\\\\
\hspace{2.5cm}
+4(1-K\cos\theta)^2(\overline{\omega}-K\sin\theta-\omega)^2=K^4\delta^2(\rho,\omega).
\end{array}
\end{equation}
In a similar way, dividing the two equations in (\ref{Ktau}) gives the equation for the time delay $\tau$ at the Hopf bifurcation as
\begin{equation}\label{Tauc}
\displaystyle{\tan[2(\theta-\omega\tau)]=\frac{2(1-K\cos\theta)(\overline{\omega}-\omega-K\sin\theta)}
{(1-K\cos\theta)^2-(\overline{\omega}-\omega-K\sin\theta)^2+\Delta^2/4}.}
\end{equation}

\noindent When the coupling phase vanishes $(\theta=0)$, the above equations for $(K,\tau)$ simplify to
\begin{equation}
\begin{array}{l}
\displaystyle{\left[(1-K)^2-(\overline{\omega}-\omega)^2+\frac{\Delta^2}{4}\right]^2+4(1-K)^2(\overline{\omega}-\omega)^2=K^4\delta^2(\rho,\omega),}\\\\
\displaystyle{\tan(2\omega\tau)=\frac{2(1-K)(\omega-\overline{\omega})}
{(1-K)^2-(\overline{\omega}-\omega)^2+\Delta^2/4}.}
\end{array}
\end{equation}

\noindent For $\Delta=0$, we have two identical oscillators with the same frequency $\omega_1=\omega_2=\omega_0$. Substituting this into equation (\ref{Freq}) gives the equation for the Hopf frequency in the form
\begin{equation}\label{Do0om}
(1-K\cos\theta)^2+(\overline{\omega}-\omega-K\sin\theta)^2=K^2\delta(\rho,\omega),
\end{equation}

\begin{figure}
\hspace{1cm}
\epsfig{file=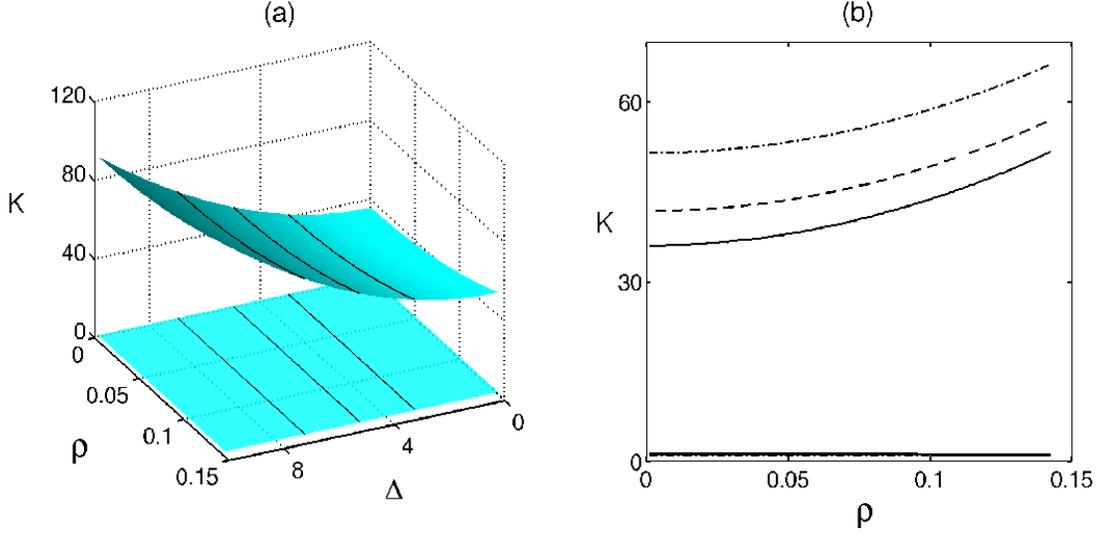,width=15cm}
\caption{(Colour online) Amplitude death region $\theta=0$, $\tau=0.15$, $\overline{\omega}=10$, is bounded by the two surfaces in figure (a). Lines in (b) show cross-sections for $\Delta=3$ (solid), $\Delta=5$ (dashed), and $\Delta=7$ (dash-dotted).}\label{uni2}
\end{figure}

\noindent and the equation for the critical time delay $\tau$ at the Hopf bifurcation reduces to
\[
\begin{array}{l}
\displaystyle{\tan[2(\theta-\omega\tau)]=\frac{2(1-K\cos\theta)(\overline{\omega}-\omega-K\sin\theta)}
{(1-K\cos\theta)^2-(\overline{\omega}-\omega-K\sin\theta)^2}=\frac{2z}{1-z^2},}\\\\
\displaystyle{z=\frac{\overline{\omega}-\omega-K\sin\theta}{1-K\cos\theta}.}
\end{array}
\]
\noindent Using the trigonometric identity $\tan2\alpha=2\tan\alpha/(1-\tan^2\alpha)$, we find
\begin{equation}\label{Do0tau}
\tan(\theta-\omega\tau)=\frac{\overline{\omega}-\omega-K\sin\theta}{1-K\cos\theta}.
\end{equation}
Equations (\ref{Do0om}) and (\ref{Do0tau}) have been recently studied by Kyrychko {\it et al.} (2011), where the effects of the width of delay distribution $\rho$, as well as coupling strength $K$ and the coupling phase $\theta$ on the amplitude death were investigated. In the case of vanishing coupling phase $(\theta=0)$, these equations reduce even further to a system studied by Atay (2003).

\newpage
\begin{figure}[h]
\hspace{0.5cm}
\epsfig{file=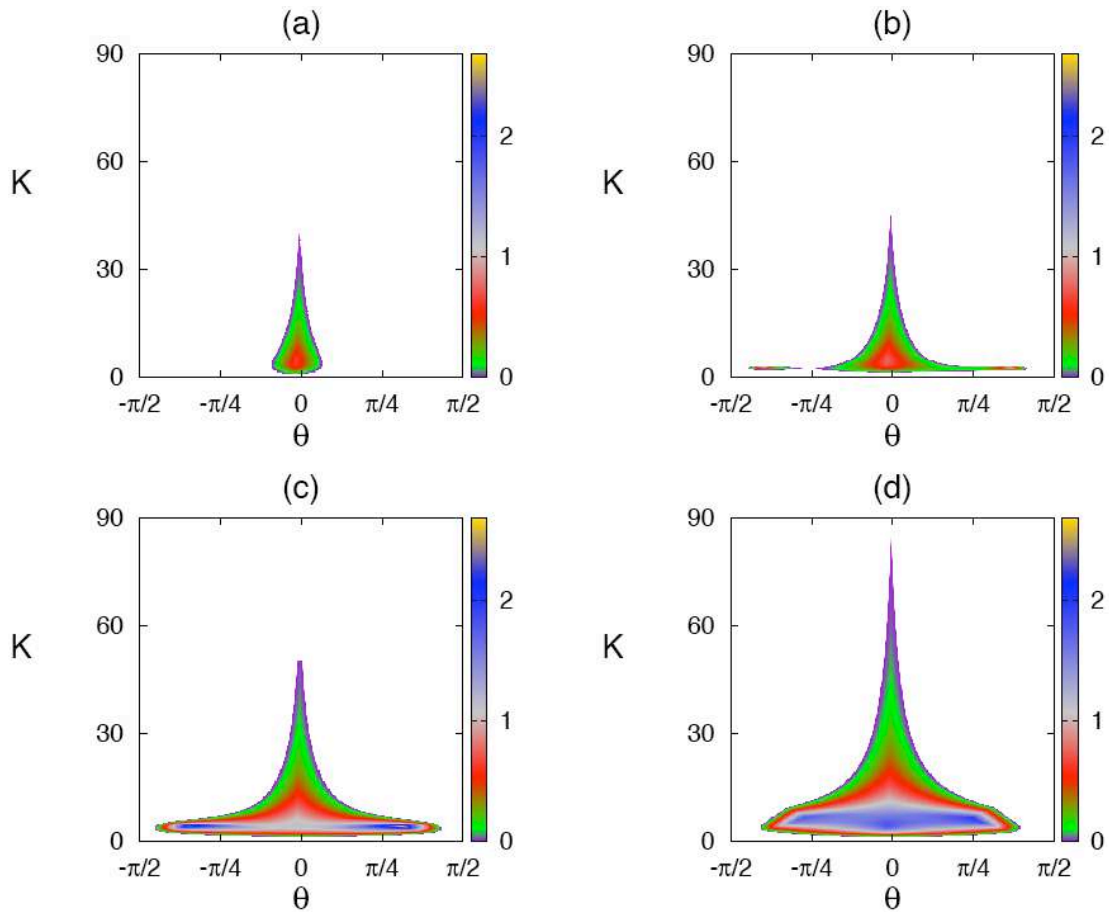,width=15cm}
\caption{(Colour online) Regimes of amplitude death depending on the coupling strength $K$ and coupling phase $\theta$ for the uniform delay distribution kernel with $\tau=0.15$, $\rho=0.002$ and $\overline{\omega}=10$. Colour code denotes $[-{\rm max}\{{\rm Re}(\lambda)\}]$ for ${\rm max}\{{\rm Re}(\lambda)\}\le 0$. (a) $\Delta=0$. (b) $\Delta=3$. (c) $\Delta=6$. (d) $\Delta=9$.}\label{uni3}
\end{figure}

\noindent To illustrate the effects of varying the coupling strength $K$ and the time delay $\tau$ on the (in)stability of the trivial steady state, we now compute the stability
 boundaries (\ref{Freq})-(\ref{Tauc}) as parameterized by the Hopf frequency $\omega$. Besides the stability boundaries themselves, which enclose the amplitude death regions, we also compute the maximum real part of the eigenvalues using the traceDDE package in Matlab.
In order to compute these eigenvalues, we introduce real variables $z_{1r,i}$ and $z_{2r,i}$, where $z_1=z_{1r}+iz_{1i}$ and $z_2=z_{2r}+iz_{2i}$, and rewrite the linearized system (SL) with the distributed kernel (\ref{UKer}) as
\begin{equation}\label{Trace}
\displaystyle{\dot{\bf z}(t)=L_0 {\bf z}(t)+\frac{K}{2\rho}\int_{-(\tau+\rho)}^{-(\tau-\rho)}M{\bf z}(t+s)ds,}
\end{equation}
where
\[
{\bf z}=(z_{1r},z_{1i},z_{2r},z_{2i})^{T},\hspace{0.3cm}
L_0=\left(
\begin{array}{ll}
N_1&{\bf 0}_{2}\\
{\bf 0}_{2}&N_2
\end{array}
\right),
\]
and also
\[
\begin{array}{c}
M=\left(
\begin{array}{ll}
{\bf 0}_{2}&R\\
R&{\bf 0}_{2}
\end{array}
\right),\hspace{0.5cm}
R=\left(
\begin{array}{ll}
\cos\theta&\sin\theta\\
-\sin\theta&\cos\theta
\end{array}
\right),\\\\
N_1=\left(
\begin{array}{ll}
1-K\cos\theta&K\sin\theta-\omega_1\\
\omega_1-K\sin\theta&1-K\cos\theta
\end{array}
\right),\hspace{0.5cm}
N_2=\left(
\begin{array}{ll}
1-K\cos\theta&K\sin\theta-\omega_2\\
\omega_2-K\sin\theta&1-K\cos\theta
\end{array}
\right),
\end{array}
\]
\begin{figure}
\hspace{0.5cm}
\epsfig{file=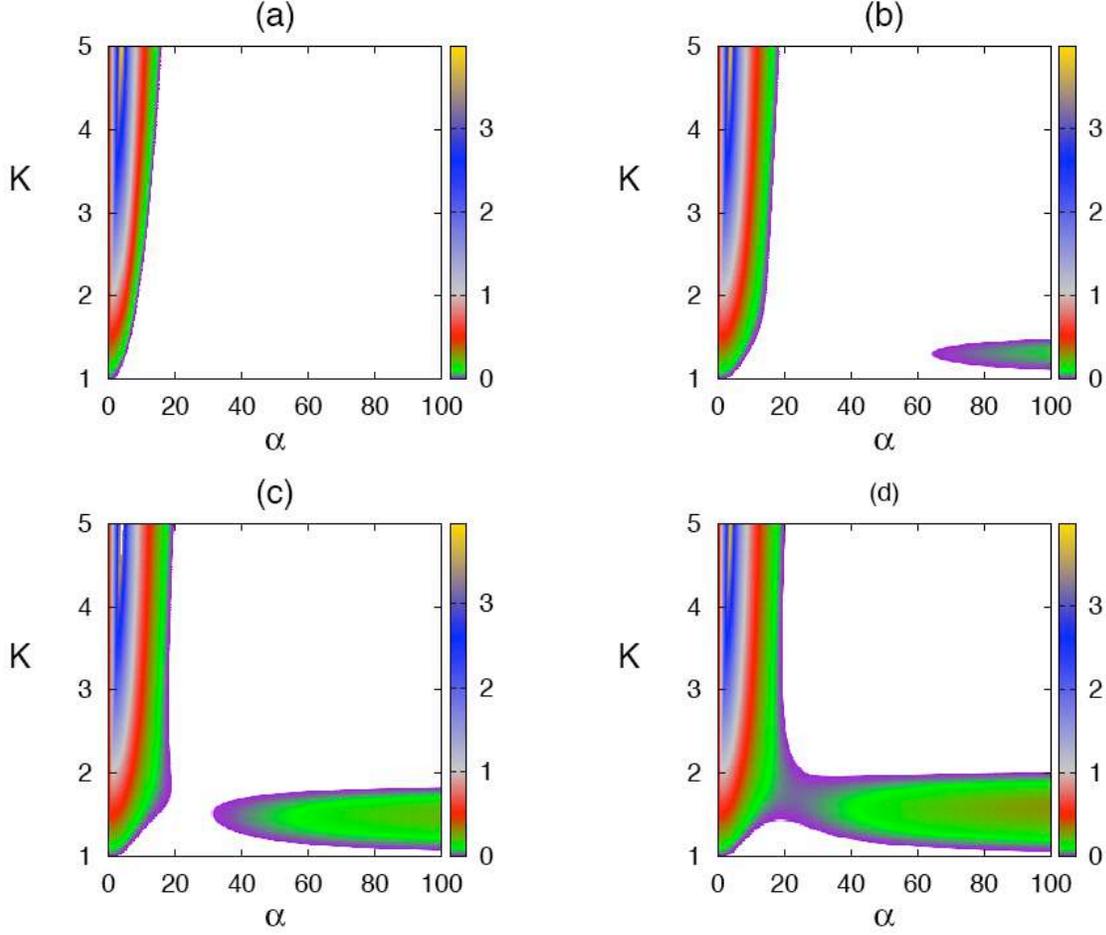,width=15cm}
\caption{(Colour online) Regimes of amplitude death depending on the coupling strength $K$ and $\alpha$ for the weak delay distribution kernel (\ref{GD}) with $\theta=0$, $p=1$ and $\overline{\omega}=10$. Colour code denotes $[-{\rm max}\{{\rm Re}(\lambda)\}]$ for ${\rm max}\{{\rm Re}(\lambda)\}\le 0$. (a) $\Delta=0$. (b) $\Delta=1.45$. (c) $\Delta=1.65$. (d) $\Delta=1.75$.}\label{p1theta0}
\end{figure}
and ${\bf 0}_{2}$ denotes a $2\times 2$ zero matrix. When $\rho=0$, the last term in the system (\ref{Trace}) turns into $KM{\bf z}(t-\tau)$, which describes the system with a single discrete time delay $\tau$. System (\ref{Trace}) is in the form in which it is amenable to the algorithms described in Breda {\it et al.} (2006) and implemented in traceDDE.

First of all, we consider the case $\rho=0$ and $\theta=0$, corresponding to a discrete time delay and vanishing coupling phase, as shown in Fig.~\ref{uni1}. This figure shows that as the frequency detuning increases, this leads to emergence of new islands of amplitude death, and for sufficiently high detuning $\Delta$, these islands merge into a single continuous region in the plane of coupling strength $K$ and average time delay $\tau$, where amplitude death can occur for an arbitrary value of $\tau$, provided $K$ lies in the appropriate range. One can note that unlike the case of identical oscillators considered

\newpage
\begin{figure}
\hspace{0.5cm}
\epsfig{file=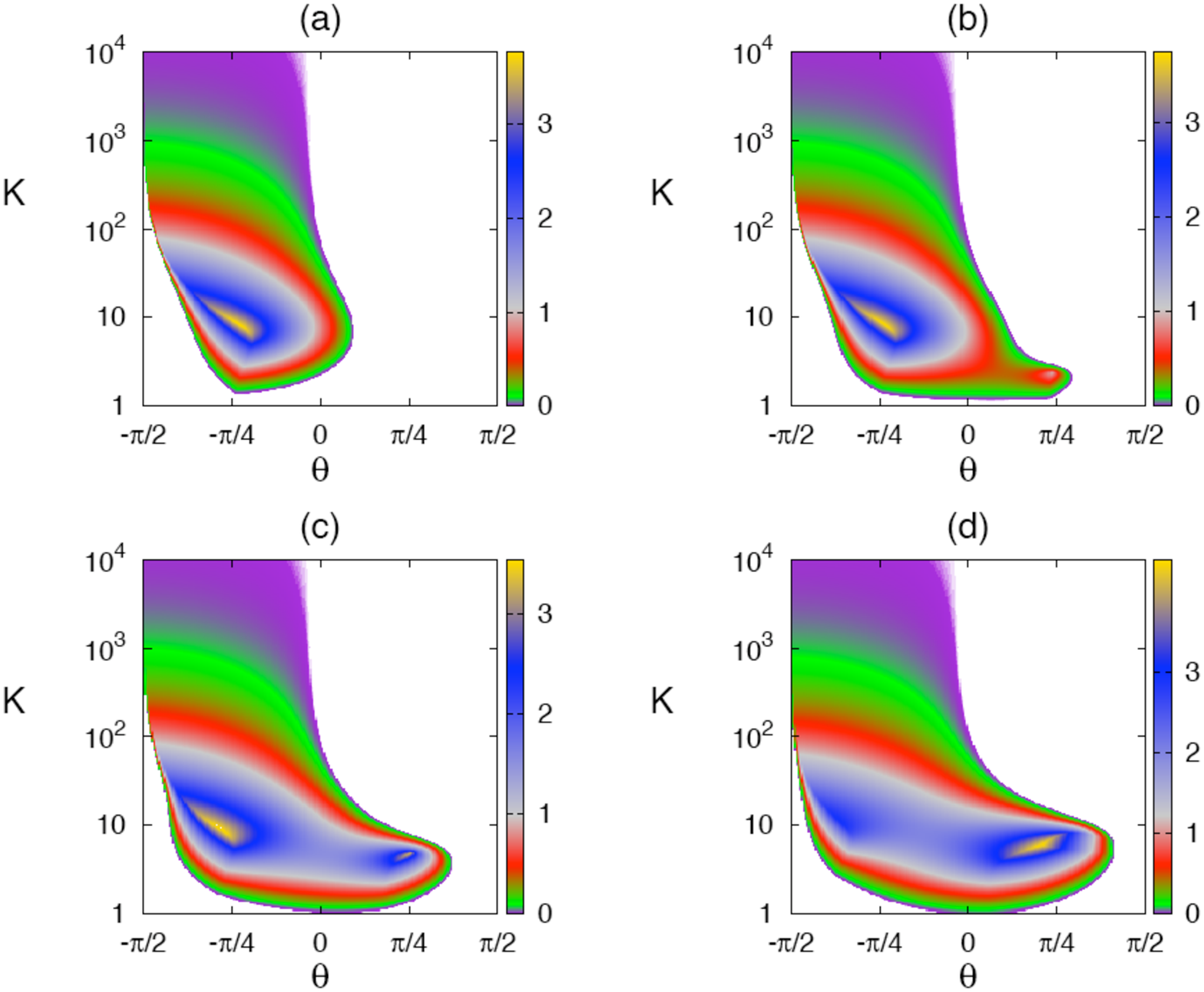,width=15cm}
\caption{(Colour online) Regimes of amplitude death depending on the coupling strength $K$ and the coupling phase $\theta$ for the weak delay distribution kernel (\ref{GD}) with $\alpha=10$, $p=1$ and $\overline{\omega}=10$. Colour code denotes $[-{\rm max}\{{\rm Re}(\lambda)\}]$ for ${\rm max}\{{\rm Re}(\lambda)\}\le 0$. (a) $\Delta=0$. (b) $\Delta=2$. (c) $\Delta=5$. (d) $\Delta=9$.}\label{p1theta}
\end{figure}
\noindent in Kyrychko {\it et al.} (2011) in this situation the values of $K$ needed to achieve stabilization of the trivial steady state for any $\tau$ are in the lower part of the overall range of $K$ values. Increase in the width of uniform delay distribution $\rho$ leads to a corresponding increase in the region of amplitude death, as illustrated in Fig.~\ref{uni2}. When we consider the impact of coupling phase on amplitude death, as shown in Fig.~\ref{uni3}, it becomes clear that the largest range of admissible $K$ values is attained for $\theta=0$, but overall the range of coupling phases, for which amplitude death is possible, increases with the frequency detuning $\Delta$. It is worth noting, however, that unlike the situation considered in Kyrychko {\it et al.} (2011) where increase in the width of distribution $\rho$ led to the asymmetry of amplitude death regions with regards to the coupling phase, as $\Delta$ is increased, the region of amplitude death remains symmetric in $\theta$ (for small $\rho$).

\begin{figure}
\epsfig{file=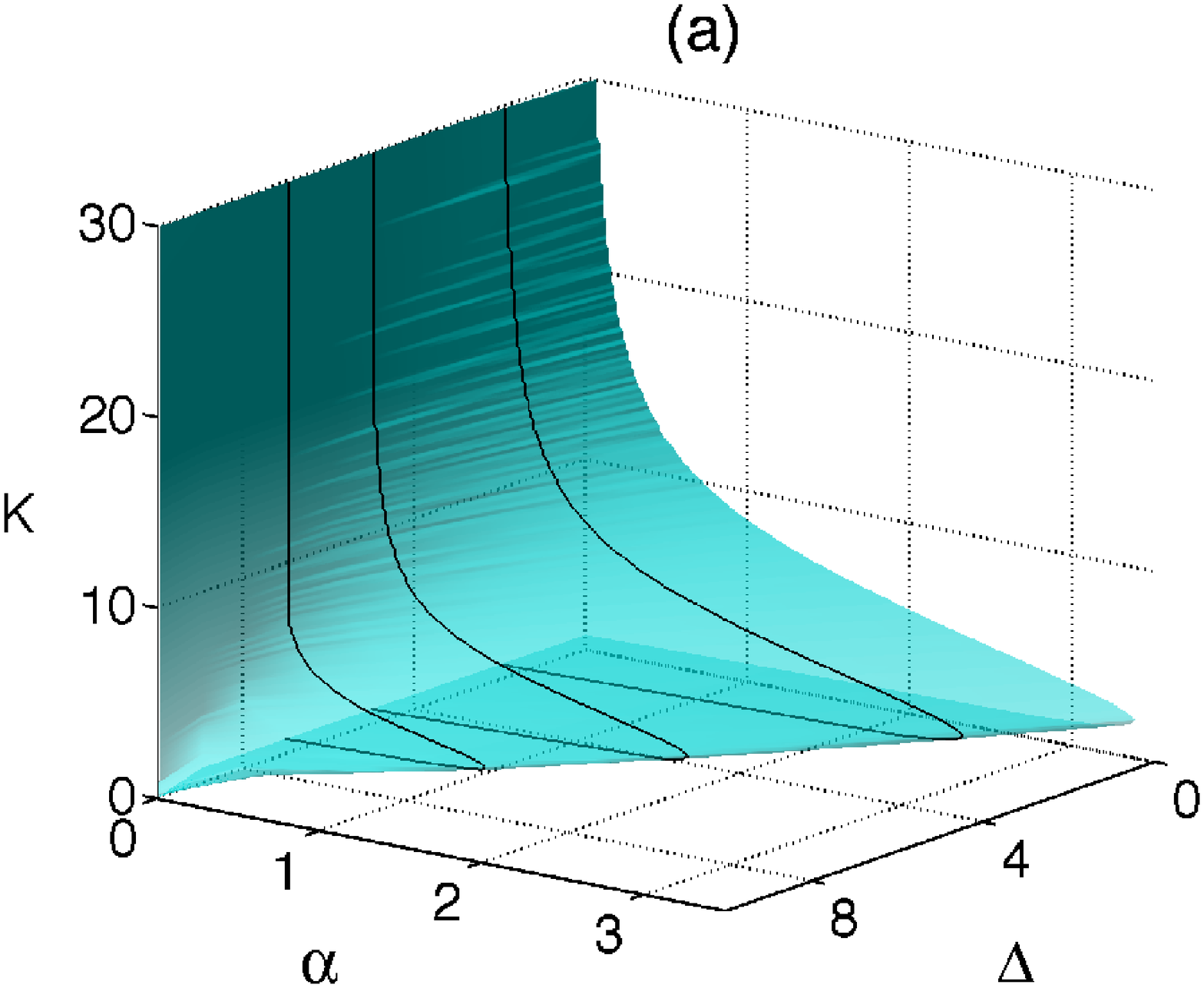,width=8cm}
\epsfig{file=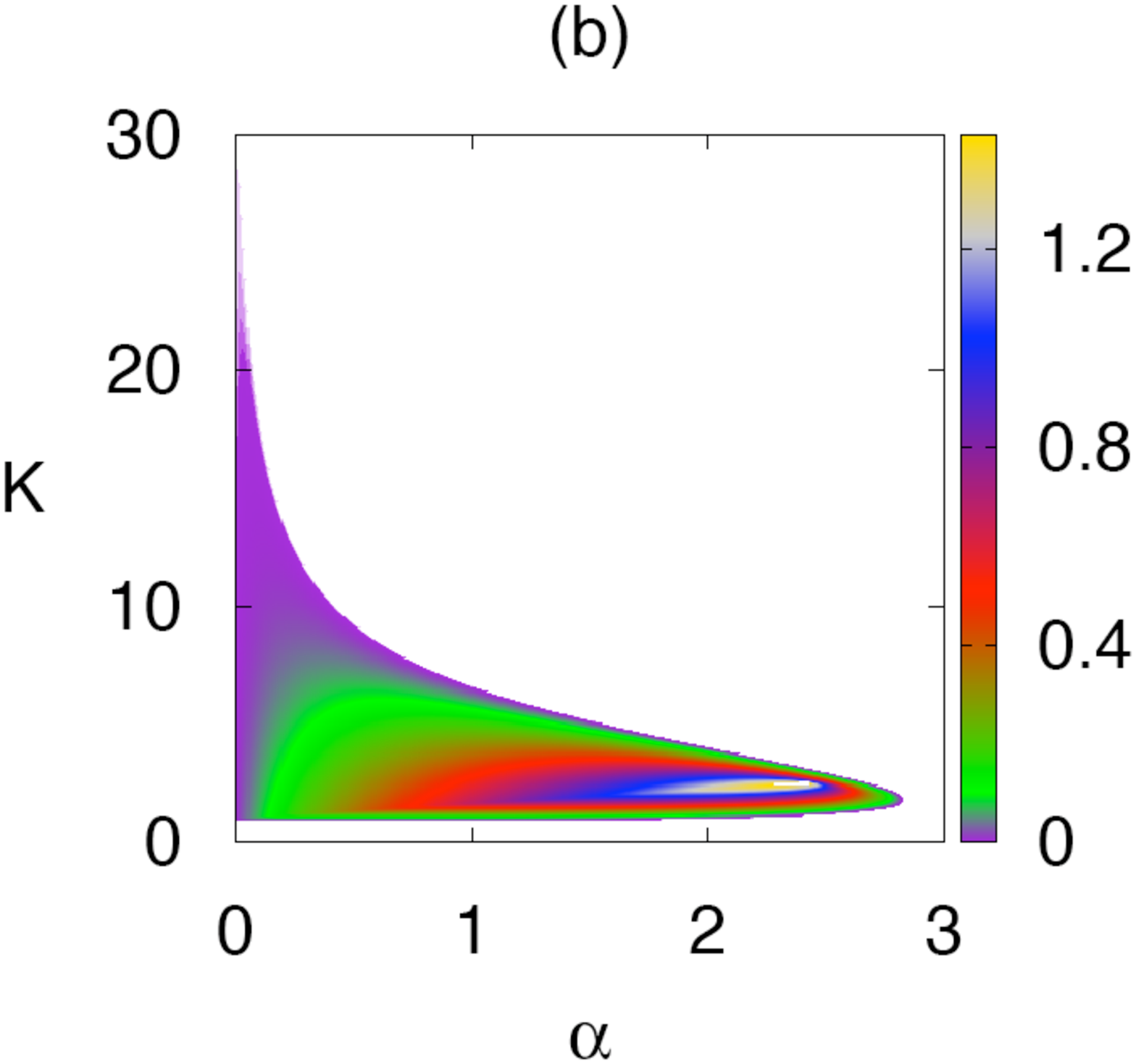,width=7cm}
\caption{(Colour online) Regimes of amplitude death depending on the coupling strength $K$ and $\alpha$ for the strong delay distribution kernel (\ref{GD}) with $\theta=0$, $p=2$ and $\overline{\omega}=10$. (a) Stability boundary. (b) Stability boundary for $\Delta=2$. Colour code denotes $[-{\rm max}\{{\rm Re}(\lambda)\}]$ for ${\rm max}\{{\rm Re}(\lambda)\}\le 0$.}\label{p2theta0}
\end{figure}

The second example we consider is that of a gamma distribution, which can be written as
\begin{equation}
g(u)=\frac{u^{p-1}\alpha^{p}e^{-\alpha u}}{\Gamma(p)},
\end{equation}
with $\alpha,p\geq 0$, and $\Gamma(p)$ being the Euler gamma function defined by $\Gamma(0)=1$ and $\Gamma(p+1)=p\Gamma(p)$.
For integer powers $p$, this can be equivalently written as
\begin{equation}\label{GD}
g(u)=\frac{u^{p-1}\alpha^{p}e^{-\alpha u}}{(p-1)!}.
\end{equation}
\noindent For $p=1$ this is simply an exponential distribution (also called a {\it weak delay kernel}) with the maximum contribution to the coupling coming from the present values of variables $z_{1}$ and $z_{2}$. For $p>1$ (known as {\it strong delay kernel} in the case $p=2$), the biggest influence on the coupling at any moment of time $t$ is from the values of $z_{1,2}$ at $t-(p-1)/\alpha$. The delay distribution (\ref{GD}) has the mean time delay
\begin{equation}\label{taum}
\displaystyle{\tau_{m}=\int_{0}^{\infty}ug(u)du=\frac{p}{\alpha},}
\end{equation}
and the variance
\[
\displaystyle{\sigma^2=\int_{0}^{\infty}(u-\tau_m)^{2}g(u)du=\frac{p}{\alpha^2}}.
\]
When studying stability of the trivial steady state of the system (\ref{SL}) with the delay distribution kernel (\ref{GD}), one could use the same strategy as the one described for uniform distribution. The Laplace transform of the distribution kernel in this case is given by
\[
\{\mathcal{L}g\}(\lambda)=\frac{\alpha^{p}}{(\lambda+\alpha)^{p}}.
\]
Substituting this into the characteristic equation (\ref{ch_eq}) yields a polynomial equation for $\lambda$:
\begin{equation}\label{Geq}
\left(1+i\omega_1 -Ke^{i\theta}-\lambda\right)\left(1+i\omega_2 -Ke^{i\theta}-\lambda\right)(\lambda+\alpha)^{2p}=K^{2}e^{2i\theta}\alpha^{2p}.
\end{equation}
In the Appendix we show how the same characteristic equation can be derived using a linear chain trick without resorting to the Laplace transform.

Figure~\ref{p1theta0} illustrates the regions of amplitude death for a weak delay distribution kernel (\ref{GD}) with $p=1$, vanishing coupling phase and increasing frequency detuning $\Delta$. Similarly to the case
\newpage
\begin{figure}[h]
\hspace{0.5cm}
\epsfig{file=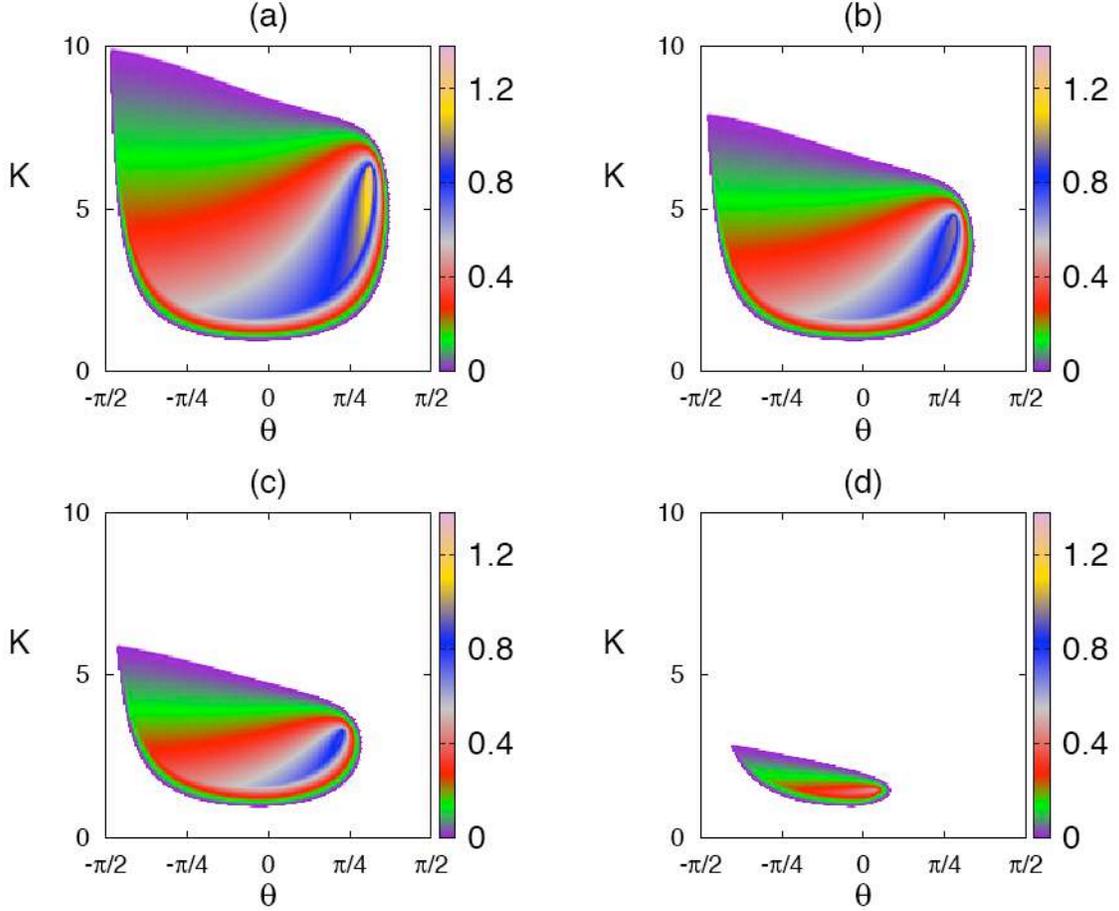,width=15cm}
\caption{(Colour online) Regimes of amplitude death depending on the coupling strength $K$ and the coupling phase $\theta$ for the strong delay distribution kernel (\ref{GD}) with $\alpha=1$, $p=2$ and $\overline{\omega}=10$. (a) $\Delta=0$. (b) $\Delta=2$. (c) $\Delta=4$. (d) $\Delta=7$. Colour code denotes $[-{\rm max}\{{\rm Re}(\lambda)\}]$ for ${\rm max}\{{\rm Re}(\lambda)\}\le 0$.}\label{p2theta}
\end{figure}
\noindent of uniform delay distribution, as $\Delta$ increases, new islands of amplitude death appear, and eventually they merge into a single continuos region of amplitude death. Since in this figure, the regions of amplitude death are plotted in terms of $\alpha$, which according to (\ref{taum}) is the inverse average time delay $\tau$, this implies that for the case of a single connected region of amplitude death in the $(\alpha,K)$ plane, amplitude death can happen for an arbitrarily small value of the average time delay, provided frequency detuning is sufficiently large. In Fig.~\ref{p1theta} we illustrate how amplitude death regions depend on the coupling phase. This figure suggests that for the same average time delay, provided the coupling phase is sufficiently negative, it is possible to achieve amplitude death for an arbitrary value of the coupling strength $K$ starting from some minimal value. The regions of amplitude death are strongly asymmetric in $\theta$, and amplitude death is possible for higher positive values of $\theta$ for larger frequency detuning $\Delta$.

When one considers a strong distribution kernel (\ref{GD}) with $p=2$ with vanishing coupling phase, there is a minimum value of the average time delay required to achieve amplitude death, as shown in Fig.~\ref{p2theta0}. This figure also suggests that the actual value of the minimum average time delay increases with the increasing frequency detuning $\Delta$. Figure~\ref{p2theta} illustrates how the coupling phase affects amplitude death. One can notice that similar to the case of a weak distribution kernel, the regions of amplitude death are asymmetric in $\theta$. However, there are two major differences from the case $p=1$, namely, that the regions of amplitude death are closed in the $(K,\theta)$ parameter plane, and these regions shrink with increasing $\Delta$ rather than grow as was the case for $p=1$. The implication is that now it is not possible to find a coupling phase, for which amplitude death could be achieved for an arbitrary value of the coupling strength $K$.

\section{Phase-locked solutions}

In the previous section we analysed the situation when the distributed time delay in the coupling leads to the destruction of stable limit cycles and stabilization of the previously unstable fixed point. In order to understand the phase dynamics of the system, we introduce new real variables $(R_1(t),R_2(t),\phi_1(t),\phi_2(t))$ as follows
\[
z_{1}(t)=R_{1}(t)e^{i\phi_{1}(t)},\hspace{0.5cm}z_{2}(t)=R_{2}(t)e^{i\phi_{2}(t)}.
\]
Here $R_{1,2}\geq 0$ and $\phi_{1,2}$ denote the amplitudes and phases of the two oscillators, respectively. Substituting this representation into the system (\ref{SL}) yields the following system of equations for the amplitude and phase variables
\begin{equation}\label{AmP}
\begin{array}{l}
\displaystyle{\dot{R}_1=\left(1-R_1^2\right)R_1+K\left[\int_{0}^{\infty}g(t')R_2(t-t')\cos[\phi_2(t-t')-\phi_1(t)+\theta]dt'-R_1\cos\theta\right],}\\\\
\displaystyle{\dot{R}_2=\left(1-R_2^2\right)R_2+K\left[\int_{0}^{\infty}g(t')R_1(t-t')\cos[\phi_1(t-t')-\phi_2(t)+\theta]dt'-R_2\cos\theta\right],}\\\\
\displaystyle{R_1\dot{\phi}_1=R_1\omega_1+K\left[\int_{0}^{\infty}g(t')R_2(t-t')\sin[\phi_2(t-t')-\phi_1(t)+\theta]dt'-R_1\sin\theta\right],}\\\\
\displaystyle{R_2\dot{\phi}_2=R_2\omega_2+K\left[\int_{0}^{\infty}g(t')R_1(t-t')\sin[\phi_1(t-t')-\phi_2(t)+\theta]dt'-R_2\sin\theta\right].}
\end{array}
\end{equation}
Significant amount of work has been done on the analysis of this system for the case of instantaneous coupling $g(s)=\delta(s)$ or discrete time delay $g(s)=\delta(s-\tau)$. In both of these cases, it has been shown that besides amplitude death, the system can exhibit a number of other interesting solutions, such as phase-locked (also known as frequency-locked) and phase drift solutions. In this section we consider effects of distributed-delay coupling on the dynamics of such solutions, which have so far remained unexplored.

\subsection{Constant amplitude dynamics}

As a first step in the analysis of system (\ref{AmP}), we assume that the amplitudes of both oscillators are equal to each other and constant: $R_1(t)=R_2(t)={\rm const}$ for all times. In this case, neglecting amplitude dynamics, the system (\ref{AmP}) reduces to a system of two {\it Kuramoto oscillators with distributed delay coupling}
\begin{equation}\label{Kur}
\begin{array}{l}
\displaystyle{\dot{\phi}_1=\omega_1+K\left[\int_{0}^{\infty}g(t')\sin[\phi_2(t-t')-\phi_1(t)+\theta]dt'-\sin\theta\right],}\\\\
\displaystyle{\dot{\phi}_2=\omega_2+K\left[\int_{0}^{\infty}g(t')\sin[\phi_1(t-t')-\phi_2(t)+\theta]dt'-\sin\theta\right].}
\end{array}
\end{equation}
It is instructive to consider this system using the variables of the phase difference $\psi=\phi_1-\phi_2$ and the mean phase $\varphi=(\phi_1+\phi_2)/2$:
\begin{equation}\label{PD}
\begin{array}{l}
\displaystyle{\dot{\varphi}=\overline{\omega}+K\left[\int_{0}^{\infty}g(t')\sin\left(\varphi(t-t')-\varphi(t)+\theta\right)\cos\frac{\psi(t)+\psi(t-t')}{2}dt'-\sin\theta\right],}\\\\
\displaystyle{\dot{\psi}=\Delta-2K\int_{0}^{\infty}g(t')\cos\left(\varphi(t-t')-\varphi(t)+\theta\right)\sin\frac{\psi(t)+\psi(t-t')}{2}dt'.}
\end{array}
\end{equation}
The last equation suggests that when $\Delta>2K$, this system exhibits phase drift (i.e. unbounded growth of the difference between the phases of two oscillators) independently of the delay distribution kernel, and hence phase locking can only occur for $\Delta<2K$. Following Schuster and Wagner (1989), we look for solutions of the system (\ref{PD}) in the form
\begin{equation}\label{PLS}
(\varphi^*,\psi^*)=(\Omega t+{\rm const},\beta),
\end{equation}
where $\Omega$ is the ensemble frequency of oscillations, and $\beta$ is the constant phase shift between the two oscillators. Substituting this into the system (\ref{PD}) yields the value of the phase shift as
\begin{equation}\label{beta_val}
\beta=\left[
\begin{array}{l}
\displaystyle{\arcsin\left[\frac{\Delta}{2K} F_c(-\Omega,\theta)^{-1} \right],}\\\\
\displaystyle{\pi-\arcsin\left[\frac{\Delta}{2K} F_c(-\Omega,\theta)^{-1} \right],}
\end{array}
\right.
\end{equation}
\noindent where we have introduced the notation
\[
F_c(a,b)= \int_{0}^{\infty}g(t')\cos(at'+b)dt',\hspace{0.5cm}F_s(a,b)= \int_{0}^{\infty}g(t')\sin(at'+b)dt'.
\]

\noindent The new common frequency $\Omega$ satisfies the transcendental equation
\begin{equation}\label{FOM}
\displaystyle{f_{\pm}(\Omega)=\overline{\omega}-\Omega-K\sin\theta\pm \frac{F_s(-\Omega,\theta)}{F_c(-\Omega,\theta)}\sqrt{K^2F_c^2(-\Omega,\theta)-\frac{\Delta^2}{4}}}=0,
\end{equation}
where the plus and minus sign correspond to the first and second value of $\beta$ in (\ref{beta_val}), respectively. It follows from this equation that possible solutions for $\Omega$ can
only lie in the admissible range
\[
F_c(-\Omega,\theta)^2\geq \frac{\Delta^2}{4K^2}.
\]
\noindent Stability of the phase-locked solution (\ref{PLS}) is determined by the roots of the corresponding characteristic equation
\begin{equation}
\begin{array}{l}
\lambda^2+2K\cos\beta F_{c}(-\Omega,\theta)+K^2[F_c(-\Omega,\theta-\beta)F_c(-\Omega,\theta+\beta)\\\\
\hspace{4cm}-F_c^L(-\Omega,\theta-\beta,\lambda)F_c(-\Omega,\theta+\beta,\lambda)]=0,
\end{array}
\end{equation}
where
\[
F_c^L(a,b,z)=\int_{0}^{\infty}g(t')\cos(at'+b)e^{-z t'}dt',
\]
\newpage
\begin{figure}[h]
\hspace{0.5cm}
\epsfig{file=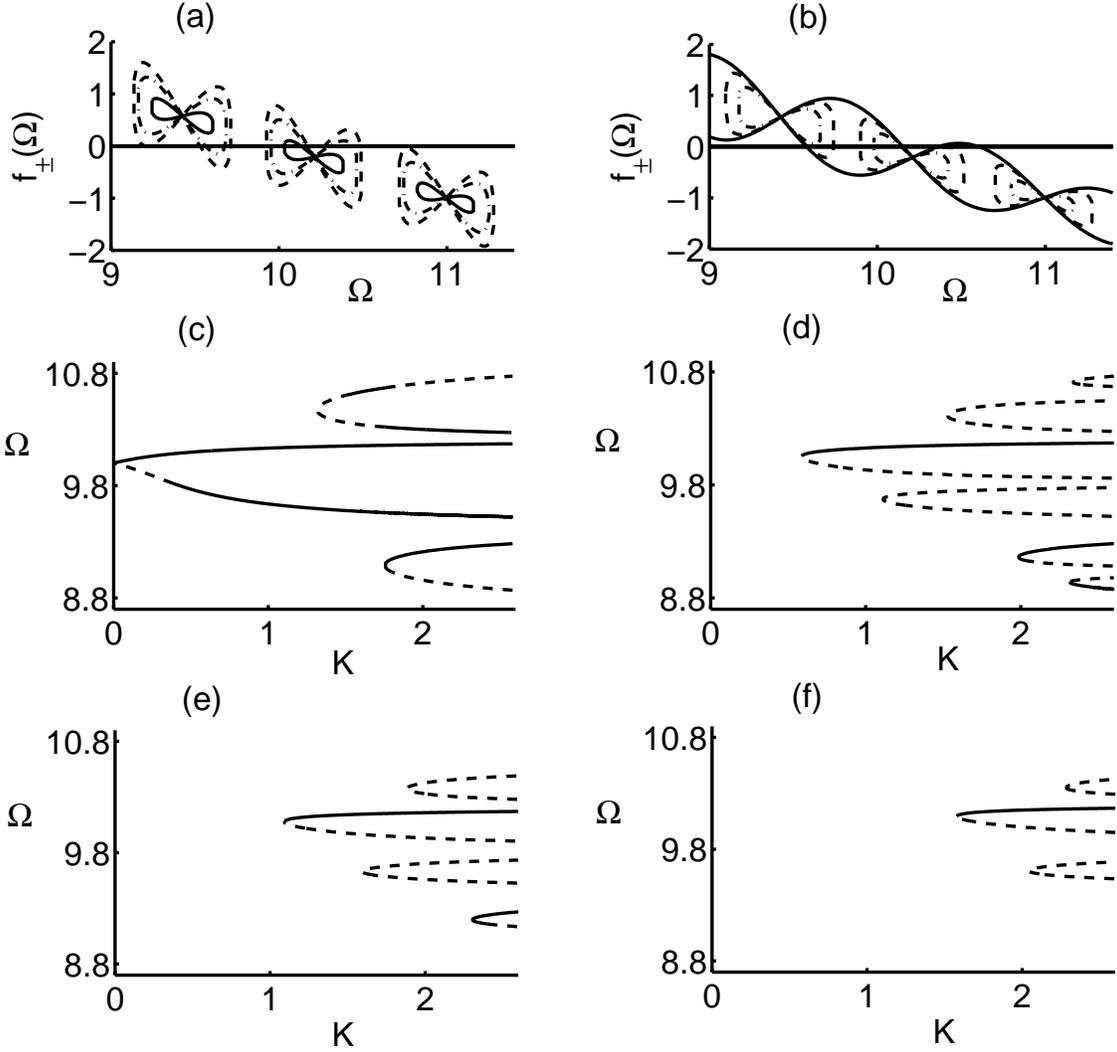,width=16cm}
\caption{(a) Function $f_{\pm}(\Omega)$ from (\ref{FOM}) for uniform distribution kernel (\ref{UKer}) with $\overline{\omega}=10$, $\tau=4$, $\rho=0.1$, 
$\theta=0$, $\Delta=1$, and $K=0.8$ (solid), $K=1.2$ (dashed), $K=1.5$ (dash-dotted). (b) Function $f_{\pm}(\Omega)$ from (\ref{FOM}) for uniform distribution kernel (\ref{UKer}) with $\overline{\omega}=10$, $\tau=4$, 
$\rho=0.2$, $\theta=0$, $K=1.5$ and $\Delta=0$ (solid), $\Delta=0.4$ (dashed), $\Delta=0.8$ (dash-dotted). (c-f) Branches of phase-locked solutions (\ref{PLS}) for uniform delay kernel with $\overline{\omega}=10$, $\rho=0.2$, $\theta=0$, $\tau=4$, and $\Delta=0$ (c), $\Delta=0.4$ (d), $\Delta=0.8$ (e), $\Delta=1.2$. Solid lines denote stable branches, dashed lines denote unstable branches.}\label{uni_branches}
\end{figure}
\noindent and the superscript $L$ refers to this integral representing the Laplace transform of a modified kernel $g(s)\cos(as+b)$.

To make further progress, we have to use specific form of a distributed kernel. For the uniform distribution (\ref{UKer}), functions $F_c$ and $F_s$ can be computed as follows
\[
F_c(a,b)=\frac{1}{\rho a}\sin(a\rho)\cos(a\tau+b),\hspace{0.5cm}F_s(a,b)=\frac{1}{\rho a}\sin(a\rho)\sin(a\tau+b),
\]

\noindent and the case of a discrete time delay can be recovered by setting $\rho=0$. For the weak distribution kernel (\ref{GD}) with $p=1$, we have
\[
F_c(a,b)=\alpha\frac{\alpha\cos b-a\sin b}{\alpha^2+a^2},\hspace{0.5cm}F_s(a,b)=\alpha\frac{\alpha\sin b+a\cos b}{\alpha^2+a^2},
\]
and similarly, for the strong delay kernel (\ref{GD}) with $p=2$:
\[
F_c(a,b)=\alpha^2\frac{(\alpha^2-a^2)\cos b-2\alpha a\sin b}{(\alpha^2+a^2)^2},\hspace{0.5cm}F_s(a,b)=\alpha^2\frac{(\alpha^2-a^2)\sin b+2\alpha a\cos b}{(\alpha^2+a^2)^2}.
\]

\begin{figure}
\hspace{0.5cm}
\epsfig{file=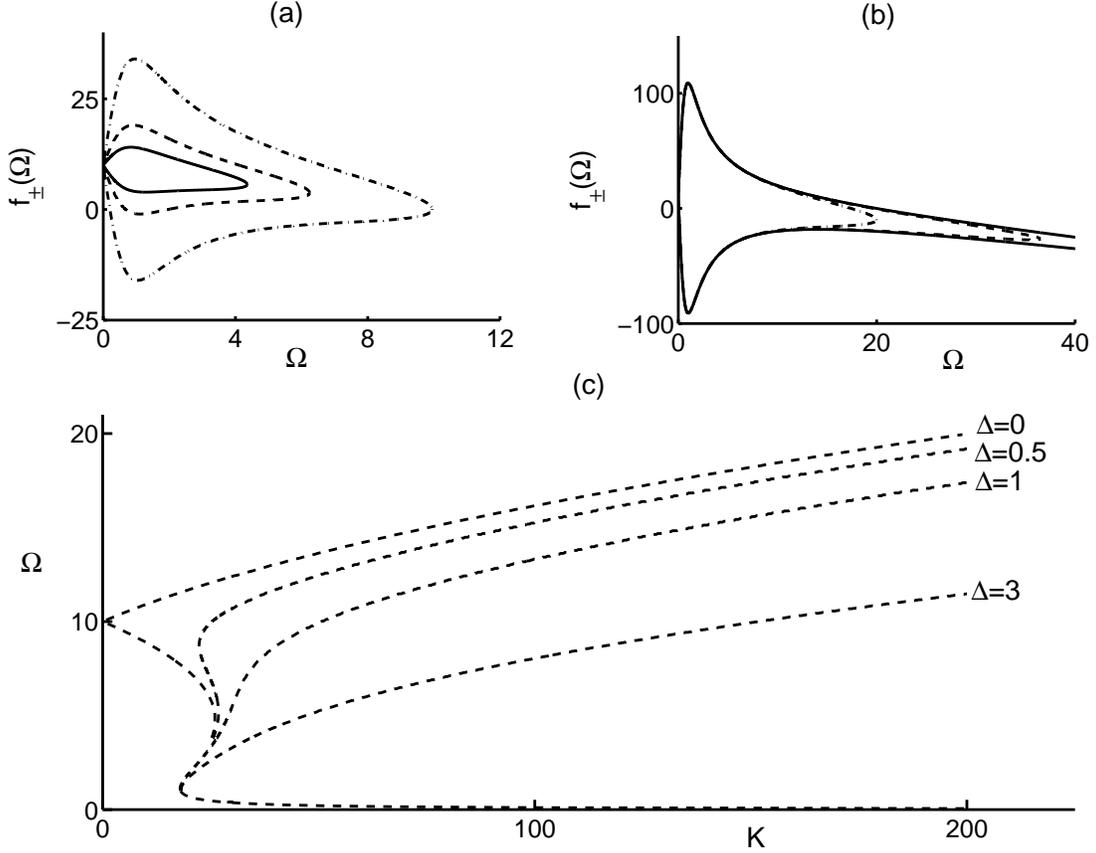,width=16cm}
\caption{(a) Function $f_{\pm}(\Omega)$ from (\ref{FOM}) for the weak distribution kernel (\ref{GD}) with $p=1$, $\overline{\omega}=10$, $\alpha=1$, $\theta=0$, $\Delta=1$ and $K=10$ (solid), $K=20$ (dashed), $K=50$ (dash-dotted). (b) Function $f_{\pm}(\Omega)$ from (\ref{FOM}) for the weak distribution kernel (\ref{GD}) with $p=1$, $\overline{\omega}=10$, $\alpha=1$, $\theta=0$, $K=200$ and $\Delta=0$ (solid), $\Delta=0.4$ (dashed), $\Delta=0.8$ (dash-dotted). (c) Branches of phase-locked solutions (\ref{PLS}) for the weak delay distribution kernel (\ref{GD}) with $\overline{\omega}=10$, $p=1$, and $\alpha=1$. All branches are unstable.}\label{p1_branches}
\end{figure}

Figure~\ref{uni_branches} illustrates how the function $f_{\pm}(\Omega)$, whose roots determine the values of the collective frequency $\Omega$ of phase-locked branches, depends on frequency detuning $\Delta$ and the coupling strength $K$. Computation of the stability of the branches shown in this Figure indicates that as the frequency detuning increases, branches of phase-locked solutions start to appear for higher values of the coupling strength. 

\newpage
\begin{figure}[h]
\hspace{0.5cm}
\epsfig{file=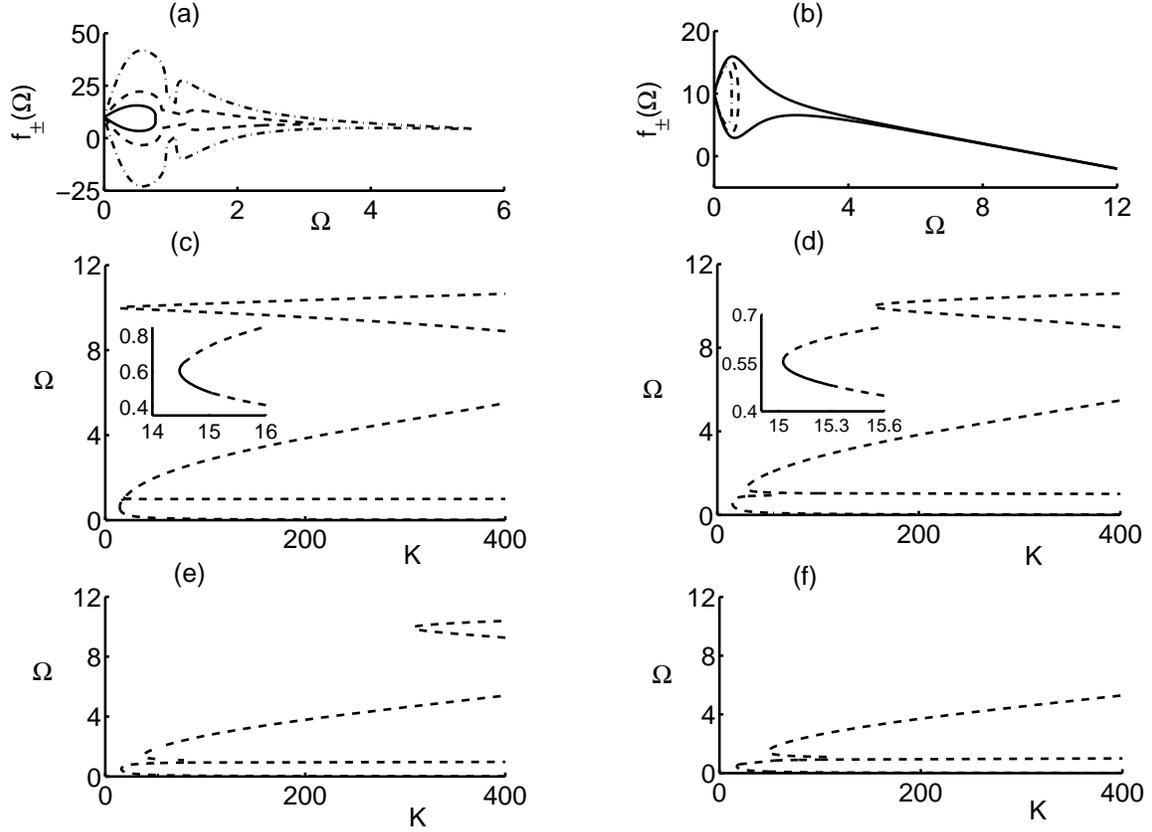,width=16cm}
\caption{(a) Function $f_{\pm}(\Omega)$ from (\ref{FOM}) for the strong distribution kernel (\ref{GD}) with $p=2$, $\overline{\omega}=10$, $\alpha=1$, $\theta=0$, $\Delta=3$ and $K=10$ (solid), $K=20$ (dashed), $K=50$ (dash-dotted). (b) Function $f_{\pm}(\Omega)$ from (\ref{FOM}) for the strong distribution kernel (\ref{GD}) with $p=2$, $\overline{\omega}=10$, $\alpha=1$, $\theta=0$, $K=10$ and $\Delta=0$ (solid), $\Delta=4$ (dashed), $\Delta=8$ (dash-dotted). (c-f) Branches of phase-locked solutions (\ref{PLS}) for the strong delay distribution kernel (\ref{GD}) with $\overline{\omega}=10$, $p=2$, $\alpha=1$, $\theta=0$, and $\Delta=0$ (c), $\Delta=3$ (d), $\Delta=6$ (e), $\Delta=9$. Solid lines denote stable branches, dashed lines denote unstable branches.}\label{p2_branches}
\end{figure}

Similar computation for the weak delay distribution kernel, shown in Fig.~\ref{p1_branches}, suggests that in this case all branches of phase-locked solutions are unstable for any value of $\Delta$. In the case of the strong distribution kernel, which is illustrated in Fig.~\ref{p2_branches}, the branches of phase-locked solutions are stable for a very narrow range of $K$ near the lowest tip of the branch for sufficiently small detuning (see insets). As the detuning increases, this makes the branches of phase-locked solutions to appear for higher values of the coupling strength in a manner similar to the case of uniform delay distribution kernel, and all these branches are unstable.

\subsection{General phase-locked solutions}

In a more general case, when $R_1$ and $R_2$ are not required to be equal to each other, one can still look for phase-locked solutions of the system (\ref{AmP}) in the form
\[
(R_1(t),R_2(t),\phi_1(t),\phi_2(t))=(\widehat{R}_1,\widehat{R}_2,\Omega t+\beta/2,\Omega t-\beta/2),
\]
where $\widehat{R}_{1,2}$ are unknown constants, $\Omega$ is the new common frequency of oscillations, and $\beta$ is the phase shift between the two oscillators. These solutions can be found as the roots of the following

\newpage
\begin{figure}[h]
\hspace{0.5cm}
\epsfig{file=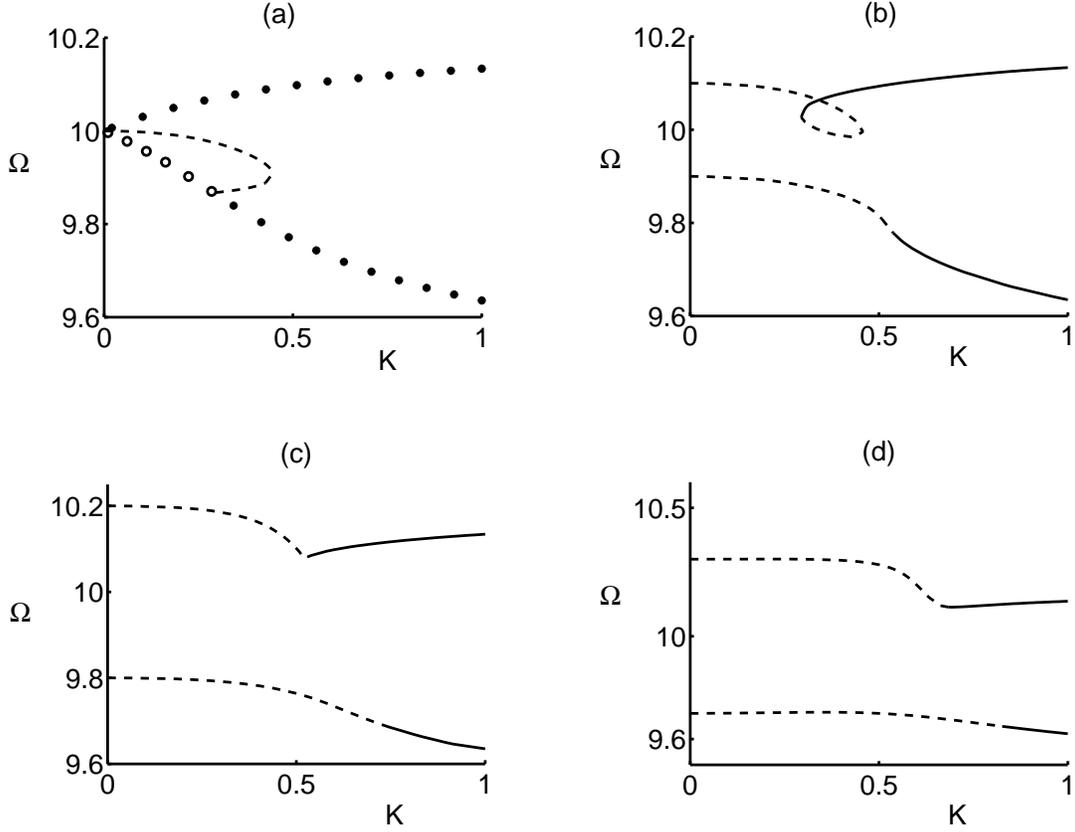,width=15cm}
\caption{Phase-locked solutions (\ref{PL_sol}) for the uniform delay distribution kernel with $\rho=0.2$, $\overline{\omega}=10$, $\tau=4$, $\theta=0$. (a) $\Delta=0$. (b) $\Delta=0.2$ (c), $\Delta=0.4$ (d), $\Delta=0.6$. Circles denote the branch with $R_1=R_2$, lines denote the branches with $R_1\neq R_2$. Solid lines (closed circles) denote stable branches, dashed lines (open circles) denote unstable branches.}\label{uni_locked}
\end{figure}
\noindent  system of equations
\begin{equation}\label{PL_sol}
\begin{array}{l}
\displaystyle{\left(1-\widehat{R}_1^2\right)\widehat{R}_1+K\left[\widehat{R}_2\int_{0}^{\infty}g(t')\cos(-\beta-\Omega t'+\theta)dt'-\widehat{R}_1\cos\theta\right]=0,}\\\\
\displaystyle{\left(1-\widehat{R}_2^2\right)\widehat{R}_2+K\left[\widehat{R}_1\int_{0}^{\infty}g(t')\cos(\beta-\Omega t'+\theta)dt'-\widehat{R}_2\cos\theta\right]=0,}\\\\
\displaystyle{\widehat{R}_1\Omega=\widehat{R}_1\omega_1+K\left[\widehat{R}_2\int_{0}^{\infty}g(t')\sin(-\beta-\Omega t'+\theta)dt'-\widehat{R}_1\sin\theta\right],}\\\\
\displaystyle{\widehat{R}_2\Omega=\widehat{R}_2\omega_2+K\left[\widehat{R}_1\int_{0}^{\infty}g(t')\sin(\beta-\Omega t'+\theta)dt'-\widehat{R}_2\sin\theta\right].}
\end{array}
\end{equation}

\noindent It is easy to check that the constant amplitude phase-locked solutions (\ref{PLS}) are only the solutions of the full system (\ref{PL_sol}) for the case of identical oscillators $\omega_1=\omega_2$, which implies $\Delta=0$ and the phase shift $\beta$ of either zero or $\pi$ (describing in-phase or anti-phase solutions, respectively).

Although it is not possible to find solutions of the system (\ref{PL_sol}) analytically, the phase-locked solutions can be computed numerically for each particular choice of delay kernel. Figure~\ref{uni_locked} shows

\newpage
\begin{figure}
\hspace{0.5cm}
\epsfig{file=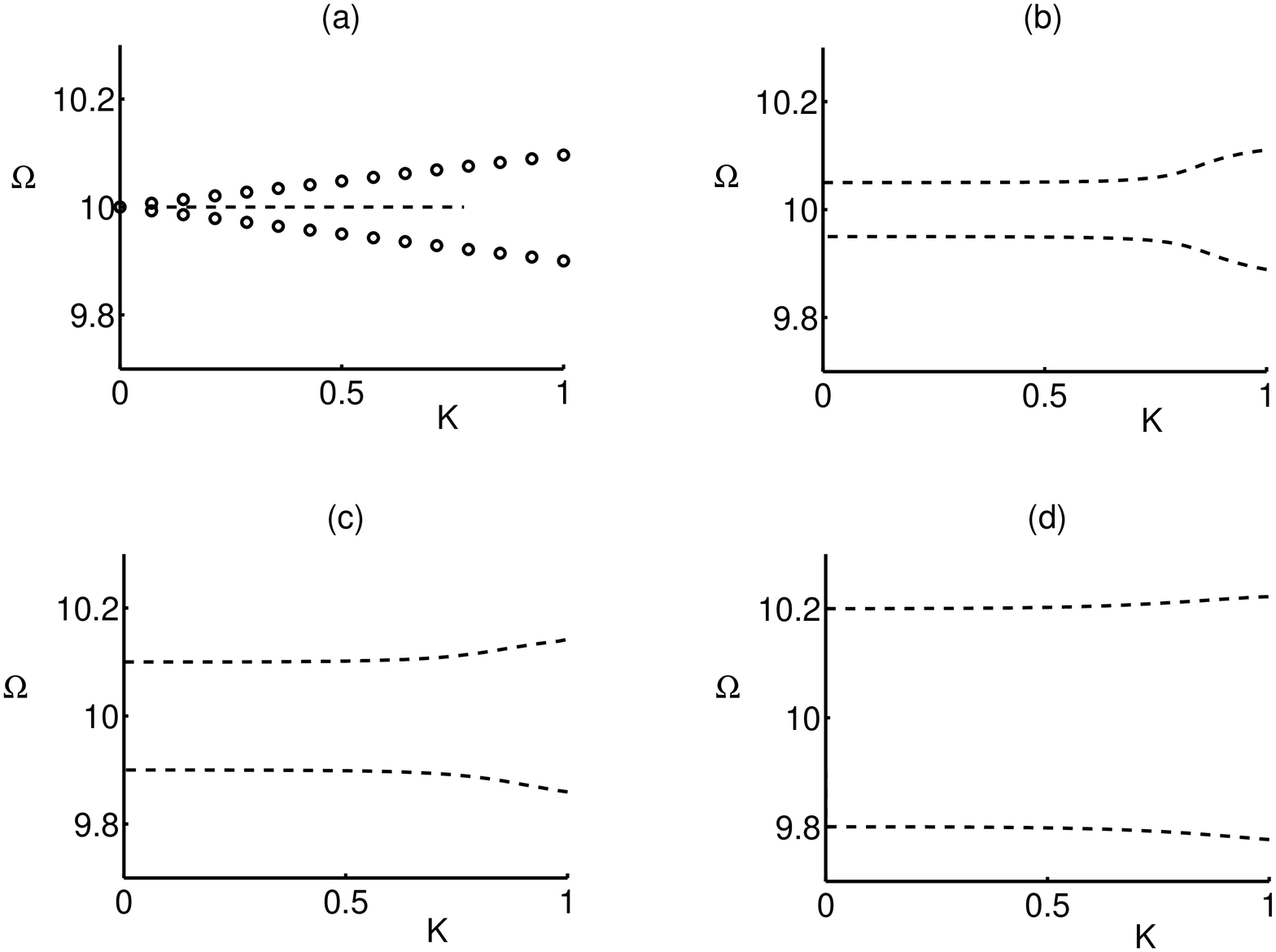,width=15cm}
\caption{Phase-locked solutions (\ref{PL_sol}) for the weak delay distribution kernel (\ref{GD}) with $\alpha=1$, $p=1$, $\overline{\omega}=10$, $\theta=0$. (a) $\Delta=0$. (b) $\Delta=0.1$ (c), $\Delta=0.2$ (d), $\Delta=0.4$. Circles denots the branch with $R_1=R_2$, black lines denote the branches with $R_1\neq R_2$. Solid lines (closed circles) denote stable branches, dashed lines (open circles) denote unstable branches.}\label{p1_locked}
\end{figure}

\noindent the branches of phase-locked solutions (\ref{PL_sol}) for the uniform delay distribution kernel. For $\Delta=0$, we identify both the branch of equal amplitude phase-locked solution, and also an unstable branch with unequal amplitudes. As the frequency detuning increases, stable branches of phase-locked solutions are observed for higher values of the coupling strength. For all values of detuning, as $K$ tends to zero, the ensemble frequency of branches of phase-locked solutions approach the values of $\Omega=\omega_{1,2}$, which should be expected from the fact that for $K=0$ the system (\ref{PL_sol}) admits solutions $(R_1,R_2,\Omega)=(1,0,\omega_1)$ and $(R_1,R_2,\Omega)=(0,1,\omega_2)$.

\begin{figure}[h]
\hspace{0.5cm}
\epsfig{file=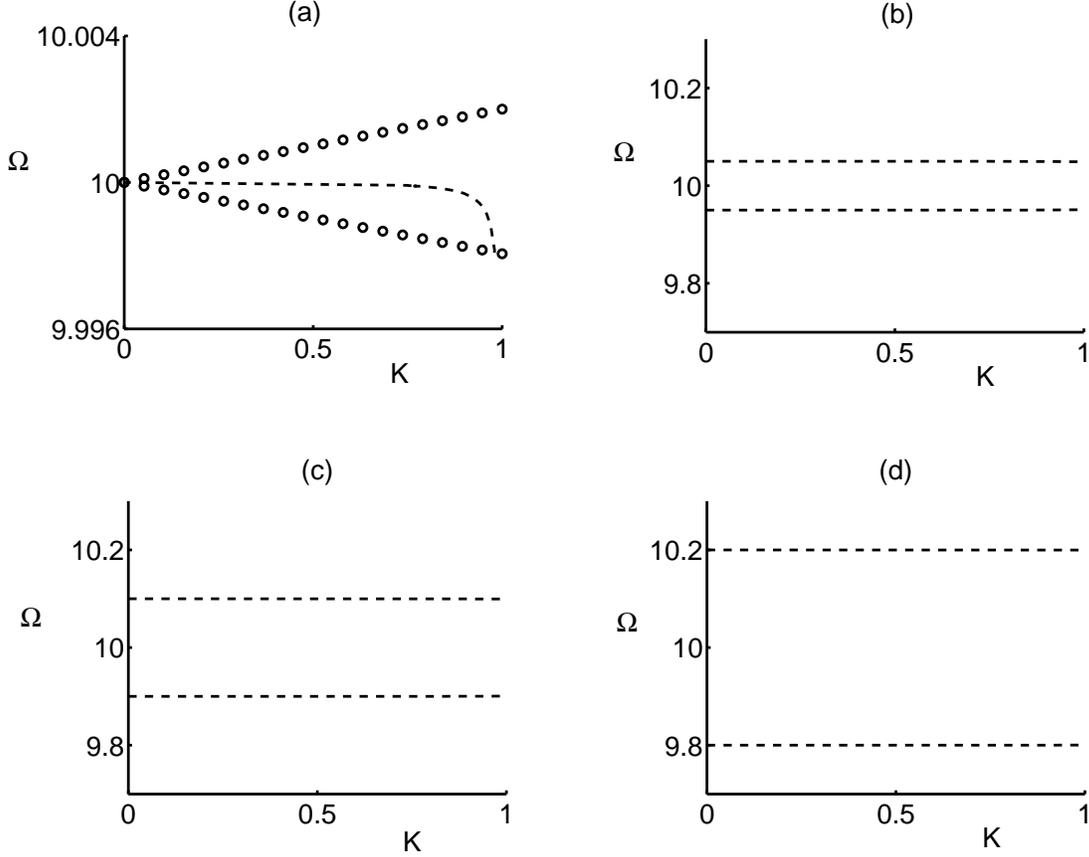,width=15cm}
\caption{Phase-locked solutions (\ref{PL_sol}) for the strong delay distribution kernel (\ref{GD}) with $\alpha=1$, $p=2$, $\overline{\omega}=10$, $\theta=0$. (a) $\Delta=0$. (b) $\Delta=0.1$ (c), $\Delta=0.2$ (d), $\Delta=0.4$. Circles denote the branch with $R_1=R_2$, black lines denote the branches with $R_1\neq R_2$. Solid lines (closed circles) denote stable branches, dashed lines (open circles) denote unstable branches.}\label{p2_locked}
\end{figure}

Figure~\ref{p1_locked} illustrates the branches of phase-locked solutions 
for the weak distribution kernel. Similarly to the case of uniform delay 
distribution kernel, for $\Delta=0$ we find the branches of solutions with 
both equal and different amplitudes, and both of these branches are unstable. For higher values of frequency detuning, we have two unstable branches of phase-locked solutions located symmetrically around the average frequency $\overline{\omega}$. The distance between these branches in terms of their ensemble frequency is increasing with increasing $\Delta$. For the strong distribution kernel, the situation is similar, as shown in Fig.~\ref{p2_locked}. Once again, we observe two unstable branches of phase-locked solutions, whose frequencies are centred around $\overline{\omega}$, and the distance between the branches grows with $\Delta$.

\newpage

\section{Discussion}

In this paper we have analysed amplitude death, as well as existence and stability of phase-locked oscillatory solutions in a generic model of Stuart-Landau oscillators with delay-distributed coupling. Regions of amplitude death have been identified for uniform and gamma distributions depending on the intrinsic system parameters (average frequency and detuning), as well as coupling strength, phase and the distribution parameters. Computation of boundaries of amplitude death regions, as well as stability eigenvalues for the trivial steady state, suggests that the more disparate the oscillators are (i.e., the higher the frequency detuning is), the larger are the regions of amplitude death in the case of uniform and weak delay distributions. However, for the strong delay distribution kernel, higher detuning corresponds to a smaller region of amplitude death. Special attention has been paid to the analysis of the effects of the coupling phase on amplitude death. While for uniform delay distribution, the maximum range of coupling strengths giving amplitude death is achieved for zero phase, in the case of gamma distribution, the non-zero coupling phase increases the range of such coupling strengths.

To understand the dynamics beyond amplitude death, we have analysed the appearance and stability of phase-locked solutions, characterized by both oscillators performing oscillations with the same common frequency and having a constant phase shift. When the amplitudes of both oscillators are constant and equal to each other, phase approximation results in a system of Kuramoto oscillators with distributed-delay coupling, in which case it is possible to find the range of admissible collective frequencies and the phase shift analytically. For a uniform delay distribution, the system of coupled Kuramoto oscillators exhibits both stable and unstable branches of phase-locked solutions, which appear for higher values of the coupling strength as the frequency detuning increases. For weak 
(exponential) delay distribution all the branches of phase-locked solutions are unstable, and for the strong (gamma function) delay distribution kernel they are stable for small detuning and sufficiently small coupling strength. We have also computed numerically branches of phase-locked solutions for uniform and gamma distributions for the full system. For a uniform delay distribution, there is co-existence of stable and unstable phase-locked branches, with branches being stable for higher values of $K$ as the detuning increases. In the case of gamma distribution, there are two branches of phase-locked solutions, which are both unstable, and their frequency difference from the average frequency $\overline{\omega}$ increases with increasing $\Delta$.

So far, we have investigated phase-locked solutions in the system of Stuart-Landau oscillators with distributed-delay coupling. One possible extension of this work would be the analysis of a phase-flip transition (Prasad 2005, Karnatak {\it et al.} 2010), where in-phase and anti-phase phase-locked branches exchange stability. This phenomenon has been observed in systems with discrete time delays, and it has even been observed in the transient dynamics preceding amplitude death. Another possibility would be to consider whether coupled systems with delay-distributed coupling are able to exhibit other types of phase dynamics and synchronization. One such scenario, which is useful in laser applications, is the case when the sum of phases of the two oscillators is constant (Erneux \& Glorieux 2010, H.-J. W\"unsche {\it et al.} 2005). Phase approximation in this case results in a delayed Adler equation, and it would be both theoretically and practically important to consider possible solutions of this model for different delay distributions. Other more complex locking scenarios have been found in quantum-dot lasers under optical injection (Lingnau {\it et al.} 2012, Pausch {\it et al.} 2012).

\section*{Acknowledgements}

This work was partially supported by DFG in the framework of SFB 910: {\em Control of self-organizing nonlinear systems: Theoretical methods and
concepts of application}. YK and KB gratefully acknowledge the hospitality of the Institut f\"ur Theoretische Physik, TU Berlin, where part of this work was completed.

\section*{Appendix}

A convenient way to derive the characteristic equation in the case of gamma distributed delay kernel is to use the {\it linear chain trick} (MacDonald 1978), which allows one to replace the original equation by an equivalent system of $(p+1)$ ordinary differential equations. To illustrate this, we consider a particular case of system (\ref{SL}) with a weak delay kernel given by (\ref{GD}) with $p=1$, which is equivalent to a low-pass filter (H\"ovel and Sch\"oll 2005):
\begin{equation}\label{WK}
g_{w}(u)=\alpha e^{-\alpha u}.
\end{equation}
Introducing new variables
\[
\begin{array}{l}
\displaystyle{Y_{1}(t)=\int_{0}^{\infty}\alpha e^{-\alpha s}z_{1}(t-s)ds,}\\\\
\displaystyle{Y_{2}(t)=\int_{0}^{\infty}\alpha e^{-\alpha s}z_{2}(t-s)ds,}
\end{array}
\]
allows us to rewrite the system (\ref{SL}) as follows
\begin{eqnarray}\label{ED_sys}
\dot{z}_1(t)&=&(1+i\omega_1)z_{1}(t)-|z_1(t)|^2z_1(t)+Ke^{i\theta}\left[Y_2(t)-z_1(t)\right],\nonumber\\
\nonumber \\
\dot{z}_2(t)&=&(1+i\omega_2)z_{2}(t)-|z_2(t)|^2z_2(t)+Ke^{i\theta}\left[Y_1(t)-z_2(t)\right],\nonumber\\\\
\dot{Y}_{1}(t)&=&\alpha z_{1}(t)-\alpha Y_{1}(t),\nonumber\\
\nonumber\\
\dot{Y}_{2}(t)&=&\alpha z_{2}(t)-\alpha Y_{2}(t),\nonumber
\end{eqnarray}
where the distribution parameter $\alpha$ is related to the mean time delay as $\alpha=1/\tau_{m}$. The trivial equilibrium $z_1=z_2=0$ of the original system (\ref{SL}) corresponds to a steady state $z_1=z_2=Y_1=Y_2=0$ of the modified system (\ref{ED_sys}).
The characteristic equation for the linearization of system (\ref{ED_sys}) near this trivial steady state is given by
\begin{equation}
(\alpha+\lambda)^2 \left(1+i\omega_1-Ke^{i\theta}-\lambda\right)\left(1+i\omega_2-Ke^{i\theta}-\lambda\right)=K^2\alpha^2 e^{2i\theta},
\end{equation}
which is the same as equation (\ref{Geq}). For larger values of $p$, one would have to introduce a larger number of additional variables in a similar fashion (two additional variables for each increase of $p$ by 1).

\end{document}